# Time-activity budget of urban-adapted free-ranging dogs

**Arunita Banerjee[1] and Anindita Bhadra[1, *]**

[1]Behaviour and Ecology Lab, Department of Biological Sciences,

Indian Institute of Science Education and Research Kolkata,

Mohanpur Campus, Mohanpur 741246

West Bengal, India

[*]**Address for Correspondence:**

Behaviour and Ecology Lab, Department of Biological Sciences,

Indian Institute of Science Education and Research Kolkata

Mohanpur Campus, Mohanpur,

PIN 741246, West Bengal, India

*tel.*+91-33 6136 0000 ext 1223

*fax*+91-33-25873020

*e-mail:*abhadra@iiserkol.ac.in

**Abstract**

The domestic dog is known to have evolved from gray wolves, about 15,000 years ago. They majorly exist as free-ranging populations across the world. They are typically scavengers and well adapted to living among humans. Most canids living in and around urban habitats tend to avoid humans and show crepuscular activity peaks. In this study, we carried out a detailed population-level survey on free-ranging dogs in West Bengal, India, to understand the activity patterns of free-ranging dogs in relation to human activity. Using 5669 sightings of dogs, over a period of 1 year, covering the 24 hours of the day, we carried out an analysis of the time activity budget of free-ranging dogs to conclude that they are generalists in their habit. They remain active when humans are active. Their activity levels are affected significantly by age class and time of the day. Multivariate analysis revealed the presence of certain behavioural clusters on the basis of time of the day and energy expenditure in the behaviours. In addition, we provide a detailed ethogram of free-ranging dogs. This, to our knowledge, is the first study of this kind, which might be used to further study the eco-ethology of these dogs.

**Introduction:**

The rapid and incessant urbanization of habitats across the world has led to an increasing number of animal species being exposed to the urban habitat and thereby, to humans[1]. This often leads to human-wildlife conflict, which is the focus of much scientific research [2,3]. However, some species have adapted well to urbanization and co-exist with humans in the urban environment. Understanding the behavioural and ecological adaptations of such species can provide important insights into the management of urban ecosystems and of mitigating human-wildlife conflict. Urban-adapted species show varied levels of tolerance of humans; while some species are extremely shy of humans, others have adjusted remarkably well to human disturbance[4–6]. Studies on some bird species have found higher flight initiation distances (FID) in populations of birds that are declining due to human disturbance, while species that are tolerant of human disturbance show no population decline[7]. Several species of mammals living in urban habitats have been known to avoid human proximity by shifting their activity peaks to crepuscular and nocturnal hours[8]. These mammals predominantly include canids, who have been known to live in and around urban habitats[9–12].

Most canids like wolves, jackals, coyotes and foxes, who share the urban ecosystem with humans, tend to avoid direct physical contact or interactions with humans[8,13]. Some studies suggest that repeated and prolonged exposure to humans lead to greater tolerance of humans in canid species[4–6]. Among the various species that share the urban habitat with humans, the domestic dog (*Canis lupus familiaris*) is special due to its ability to communicate with and befriend humans. As pets, they are bred, raised and taken care of by humans and are habituated to living in human homes [14,15]. Free-ranging dogs also seek human proximity actively and are known to interact with and comprehend social cues from unknown humans[16,17]. Due to their pro-social behaviour, dogs have been the subject of a large array of studies that aim to understand the evolutionary trajectory of the dog-human relationship.

Free-ranging dogs comprise the majority of the world's dog population[18]; they, predominantly inhabit urban and rural habitats, primarily living on streets, surviving on garbage and human-provided food [19,20]. They not only depend on humans for food but also prefer to be close to humans while giving birth [16]. They are primarily scavengers, showing a high degree of flexibility in their food habits, eating anything from vegetable peels to meat [21] and also have a high degree of flexibility in their interactions with humans [4,19,22]. Humans, on the one hand, are the major source of food and shelter for the free-ranging dogs and on the other, a cause of morbidity and mortality[23]. The free-ranging dogs provide an excellent model system for understanding the various behavioural and ecological adaptations that have been instrumental in enabling dogs to live in close proximity with humans. However, the behaviour of dogs in their natural habitat is yet to be understood in detail. Though this lacuna is being covered by some recent studies, a detailed ethogram of free-ranging dogs is yet to be published.

Over the years, ethologists have been trying to decipher reasons behind actions, and thus understand 'behaviour'. This has given rise to diversifications in the field of animal behaviour studies, particularly in the directions of cognition and psychology. Behavioural studies in recent times majorly focus on rigorous analyses of behaviours, fitting behavioural models onto certain target species and less of comprehensive observations [24–26]. However, comprehensive and meticulous observations provide greater details and insights into the behaviour of the target species. This can then be used to compile ethograms which register all individual and interactive behaviours of the target organism, in different contexts [27–32]. An ethogram is the first step towards the understanding of the behaviour of a species and thus needs to be constructed for every model system in the field of animal behaviour.

In the Canidae family, well-established ethograms of the wolf (*Canis lupus*), coyote (*C. latrans*) and red fox (*Vulpes vulpes*), dholes (*Cuon alpinus*) and singing bush dog (*Canis* *hallstromi*) have been reported[27–29,32–34]. The hunting and scent marking behaviours have been the major focus of wolf ethograms. The hunting process has been modelled into a stepwise process of search, approach, watch, attack-group, attack-individual, and capture [34], whereas, the scent marking behaviours have been categorised as raised-leg urination, squat urination, defecation, and scratching [35]. A study in dholes, an endangered social canid from the forests of southern and south-eastern Asia, reports the occurrence of 33 behaviours, categorised into locomotion, scent marking, resting, social behaviour, feeding and other miscellaneous behaviours and classifies them as crepuscular animals [28]. Coyotes have been reported to show a total of 540 behavioural patterns, with exploratory behaviours accounting for the major part of the day or time activity budget of the species [32]. An all-inclusive ethogram of New Guinea Wild dogs, a form of small dingoes, has reported the occurrence of more than 250 behaviours in the species [29]. Both in order to study free-ranging dogs and to compare their behavioural patterns with other canids, there is a need for an ethogram that is accessible to all. We have carried out detailed behavioural observations of free-ranging dogs in their natural habitats in India for several years and published partial ethograms focusing on certain aspects of their behaviour, like foraging, parental care, mating, etc. [4,19,36–38].

Time-activity budgets [39–54] help to understand the broad behavioural signatures of a species with respect to the amount of time (thus effort) it invests in the display of various behaviours. This involves extensive sampling but is important for understanding the habits of a species. Activity patterns of wolves, for example, have been known to vary according to geographic area, in different study sites. Wolves have been reported to be nocturnal in Italy [55] and Minnesota[56] and active throughout the day with crepuscular peaks in Canada [57] and Poland[58], Studies have also shown that wolves tend to avoid human proximity[13,59]. Human activity has been hypothesised to shift the activity peaks of wolves to the crepuscular and nocturnal hours when human activity is less[60]. Due to their evolutionary history, the domestic dogs can be expected to do the same, but since they have adapted to living with humans, their habit is expected to be modulated by the peaks in human activities.

In India, free-ranging dogs were studied in different locations during the day, when they share 'road space' with humans (7:00 AM to 7:00 PM). It was observed that the dogs spent about 50% of their time in activity[4]. The nature of their activity in the nocturnal hours is not well understood and needs to be studied for a better understanding of the habits of dogs in a natural habitat. This is not only necessary for a complete understanding of the behavioural repertoire of the dogs, but also because humans can also be active in the nocturnal hours, and we do not yet have any study which can help us understand the chances of human-dog conflict during the night. In this paper, for the first time, we provide a detailed time activity budget of free-ranging dogs for the whole day, based on a population level study carried out in India over a period of one year. In this study, we address the question of whether dogs are primarily nocturnal, crepuscular, or more flexible in their habits, adjusting their activity patterns to survive in a human-dominated environment. The time activity budget was prepared using a detailed ethogram, which has been compiled over a period of 8.5 years, by different members of the Dog lab at IISER Kolkata, which we also share here.

**Results**

**Dog Ethogram**

The existing ethogram of the Dog Lab of IISER Kolkata, created by compiling *ad libitum* observations for 8.5 years, carried out by a large number of observers, was updated during this study by including all newly observed behaviours, editing existing definitions, or splitting previously recorded complex behaviours into finer components (See Supplementary Information Table 1). This ethogram was used to sample behaviours, over 1 year, to create a twenty-four-hour time-activity budget of free-ranging dogs.

**Demographics**

Behavioural sampling was carried out for 1 year and 5669 sightings were accumulated. There was a significantly higher proportion of adults (N=4001) than juveniles (N=1668) in the population (Test of proportion: $\chi^2$ = 957.65, df = 1, p<2.2e-16) (Figure 2a). The proportion of juveniles in the population was significantly lower than that of adults in the period July – October, and in November, while the proportions were comparable in December – March, which is the pup-rearing season (Figure 2b) (See Supplementary Table 2 for details).

**Time-activity budget**

As seen in Figure 2c, more than fifty per cent activity was seen only in a window between 0900 hours to 2230 hours (See Supplementary information Table 3 for details). In order to check if the free-ranging dogs are primarily nocturnal, we compared their activity and inactivity levels over the diurnal (0600 hours -1800 hours) and nocturnal (1800 hours – 0600 hours) periods of the day. The activity levels were significantly higher than inactivity in both the diurnal and nocturnal hours (Test of proportion; Diurnal: $\chi2$ = 93.247, df = 1, p < 2.2e-16; Nocturnal: $\chi2$ =

6.5246, df = 1, p = 0.01064; Figure 3a, 3b). The dogs were found to be significantly more active during the diurnal hours than the nocturnal ones (Goodness of Fit: 667.64, df = 1, p = 0.00; Figure 3c).

**Generalized linear mixed model analysis:**

Activity levels varied between adults and juveniles (GLMM analysis; Table 1; Figure 4a). The time of the day and age of the dog significantly affected their activity level. The interaction of these two factors was also found to have a significant effect on the activity levels. The activity in the different time periods, when further broken down into the constituent behavioural categories, showed some degree of variation (Figure 4b). Since the original dataset had a variable number of data points in the eight time-blocks, the same analysis was carried out by randomly selecting 200 data points in each three-hour block. The results obtained did not differ from those on the whole data (See Supplementary Table 4 for details).

**Principal component analysis**

The PCA revealed that the behavioural categories are clustered to some extent (Figure 5), in accordance with energy expenditure for each category of behaviour (component 1) and the time of occurrence in the day (component 2). Post hoc Pearson's test of correlation revealed significant correlations between most pairs of behavioural categories (See Supplementary Information- Table 5 and Table 6). This suggests that though the data was collected through a random sampling protocol at a population level, there are some inherent patterns in the behavioural repertoire of the free-ranging dogs.

**Discussion:**

Activity patterns of an animal tend to ensure maximum resource utilization and minimizing risk, and are influenced by a number of factors, including physiological adaptations of the animal, availability and distribution of food and disturbances caused by predators (and humans) [61]. While most urban-adapted species show a tendency to avoid humans[62,63], we do not see such a tendency in the behaviour of free-ranging dogs, and they even show a preference for denning in close proximity to humans for the purpose of giving birth [16]. Wolves and coyotes, on the other hand, restrict their diurnal movements in order to avoid humans. They regulate their activities in such a way that they can avoid human interaction, but ensure livestock hunt [60,61,64]. Although, social behaviour patterns in the family Canidae, varies from species to species, yet it shows an evolutionarily conservative trend [65] as these animals continue to maintain their association with humans.

Our study revealed that, contrary to expectations based on their evolutionary history, the free-ranging dogs are not nocturnal. More than 50% activity levels were seen to be spread out over a large part of the day, from 0900 hours to 2230 hours. This suggests that the dogs are a generalist species with respect to their activity time, since their high activity window does not fall exclusively within diurnal or nocturnal hours; they are predominantly active during the human activity hours. Activity levels were significantly affected by the time of the day and the age class of the dog; thus, adults and juveniles perform activities with different contexts and involve different energy expenditures. The time of the day may actually play a role in determining the context of activity for adults and juveniles.

Context-dependent behavioural clustering studies have been carried out on pet dogs, which indicate that interactive behaviour of pet dogs with their owners was clustered on the basis of three components, namely, anxiety, acceptance and attachment[66]. In our study, the existence of behavioural clusters was found with respect to 'time of occurrence' and 'energy expenditure'. The PCA did not resolve the different categories of activities into very tight clusters, but, the Principal Component 1 clearly drew out the active behaviours from the inactive behavioural category, thereby differentiating behaviours with respect to the energy expenditure involved in performing them. Moreover, some minor clustering could be seen along the Principal Component 2, which seemed to be primarily on the basis of the time of occurrence. Most behavioural category pairs were found to be strongly correlated, which suggested that most behaviours occur simultaneously and involve similar amounts of energy expenditure. This strengthens the position of free-ranging dogs as a truly generalist species.

This study, though first of it's kind in free-ranging dogs, is not devoid of limitations. Keeping in mind the ease of data collection primarily due to safety issues, the data was not evenly distributed across the 24 hours of the day. However, in order to check if that could have affected the results, a control GLMM analysis was carried out with 200 randomly selected data points in each three-hour block, and the results did not differ from those of the original analysis.

Free-ranging dogs depend on human kindness for their food, and in return, they primarily guard human settlements and scavenge garbage dumps. But instead of a simplistic harmonious relationship, the dog-human dyad has evolved three paradigms, which involve mutualistic and positive interactions between dogs and humans, neutral co-existence as well as dog-human conflict, manifested through aggression from dogs towards humans as well as animal abuse and culling [67]. A better understanding of the behavioural ecology of the free-ranging dogs can help to reduce dog-human conflict and device better management strategies for dogs in the regions of the world that sustain large populations of free-ranging dogs.

**Methods:**

**Sampling**

The study comprised of collecting data on free-ranging dogs in their natural habitats in various areas through instantaneous scan sampling of free-ranging dogs in urban and semi-urban habitats. The observer walked on a pre-determined route at randomly selected time points during the day, and whenever a dog was sighted, its age class, sex, behaviour at the time of sighting were noted down, along with the date, time and location of the sighting. The observer did not interact with the focal animals in any manner, but interactions observed between the focal dogs and other humans were recorded.

Sampling was carried out in eleven different locations (SI Figure 1) – IISER Kolkata campus (22.9638° N, 88.5245° E ), Gayeshpur (22.9554° N, 88.4961° E ) Harringhata (22.9605° N, 88.5674° E), Kanchrapara (22.9441° N, 88.4335° E), Kalyani (22.9751° N, 88.4345° E ), Halisahar (22.9441° N, 88.4193° E), Naihati (22.8929° N, 88.4220° E), Barrackpore (22.7674° N, 88.3883° E), Balindi (22.9740° N, 88.5382° E) , Jaguli (22.9276° N, 88.5505° E ) and Mohanpur (23.6565° N, 88.2254° E ) between July 2016 and November 2017. Each of these sampling bouts was called a census. A total of 5669 sightings were recorded during this study, covering 24 hours of the day, and three main seasons of dog behaviour – pre-mating, mating, pup-rearing.

**Data Analysis**

All behaviours were divided into active or inactive behaviours and were subsequently categorised into certain types, based on the social context of the behaviour (Figure 1). A time activity budget of free-ranging dogs over 24 hours of the day was obtained based on the proportion of times spent by dogs in different behaviours (walking, tail wagging, sleeping, barking, eating etc). The data was then divided into two blocks to calculate and compare the activity levels in the diurnal (0600 hours – 1800 hours) and nocturnal (1800 hours- 0600 hours) hours of the day.

**Statistical methods**

The percentages of adults and juveniles in the population were compared using Test of proportion, and the percentages in each month of data collection were compared using Chi-square tests. Tests of proportion were used to compare activity and inactivity levels within diurnal and nocturnal hours, and to compare the levels of activity between the diurnal and nocturnal hours. The activity levels across the day were compared between the different age classes – adults (age > 6 months) and juveniles (age ≤ 6 months) using Chi-square tests. In order to understand the factors affecting variation in activity levels and any possible interactions between these factors, a Generalized Linear Mixed Model (GLMM) analysis was carried out, with the fixed effects being time of the day and age class of the animal, random effects being time of the year and location of sighting, and the response variable being Activity/Inactivity, considering a binomial distribution of the data. In order to understand the behavioural patterns of the dogs at a more granular level, the day was divided into eight blocks of three hours each (0000-0259, 0300-0559, 0600-0859, 0900-1159, 1200-1459, 1500-1759, 1800-2059, 2100-2359), and the proportion of time spent in various sub-categories of active behaviours was estimated. A Principal Component Analysis (PCA) was carried out to analyse clusters between behavioural categories. Frequencies for each behavioural category, in each of the eight-time blocks, were calculated and this matrix was used as input for the PCA Pearson's test of correlation was used to check correlations between all possible combinations between the behavioural categories. All data analysis was carried out using in StatistiXL and R 3.3.3.

**Acknowledgements:**

The sampling of dogs and analysis of data was carried out by A. Banerjee. The two authors co-wrote the manuscript. A. Bhadra designed and supervised the work. A. Banerjee was supported by a fellowship from IISER Kolkata, and the work was funded by IISER Kolkata. The authors would like to thank all the past and present member of the Dog Lab at IISER Kolkata, who have contributed at various levels to the compilation of the ethogram. An initial ethogram comprising of approximately 100 behaviours was prepared by A. Bhadra, to which additions were made over a period of 9 years to arrive at the ethogram reported in this manuscript.

**Competing interests**

The authors declare that there are no competing interests

**Figures:**

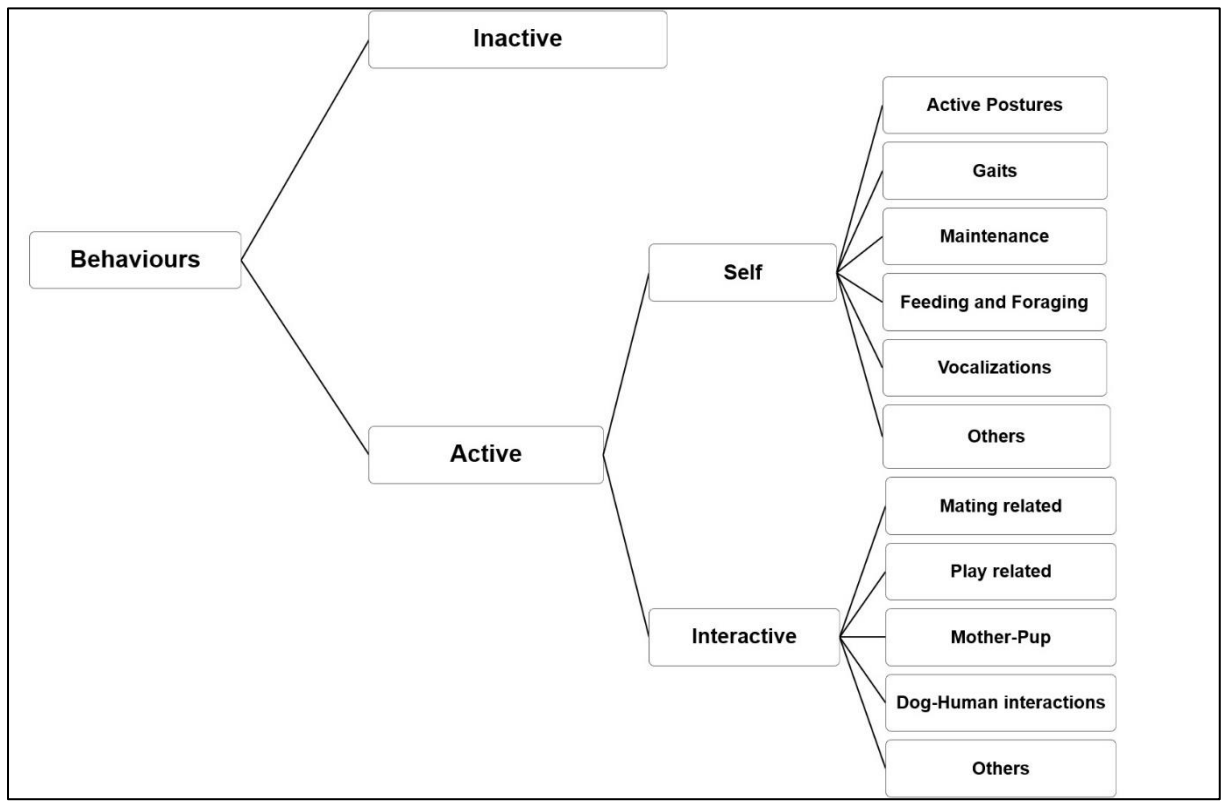

**Figure 1:** The categorisation of behaviours based on activity level and context of occurrence.

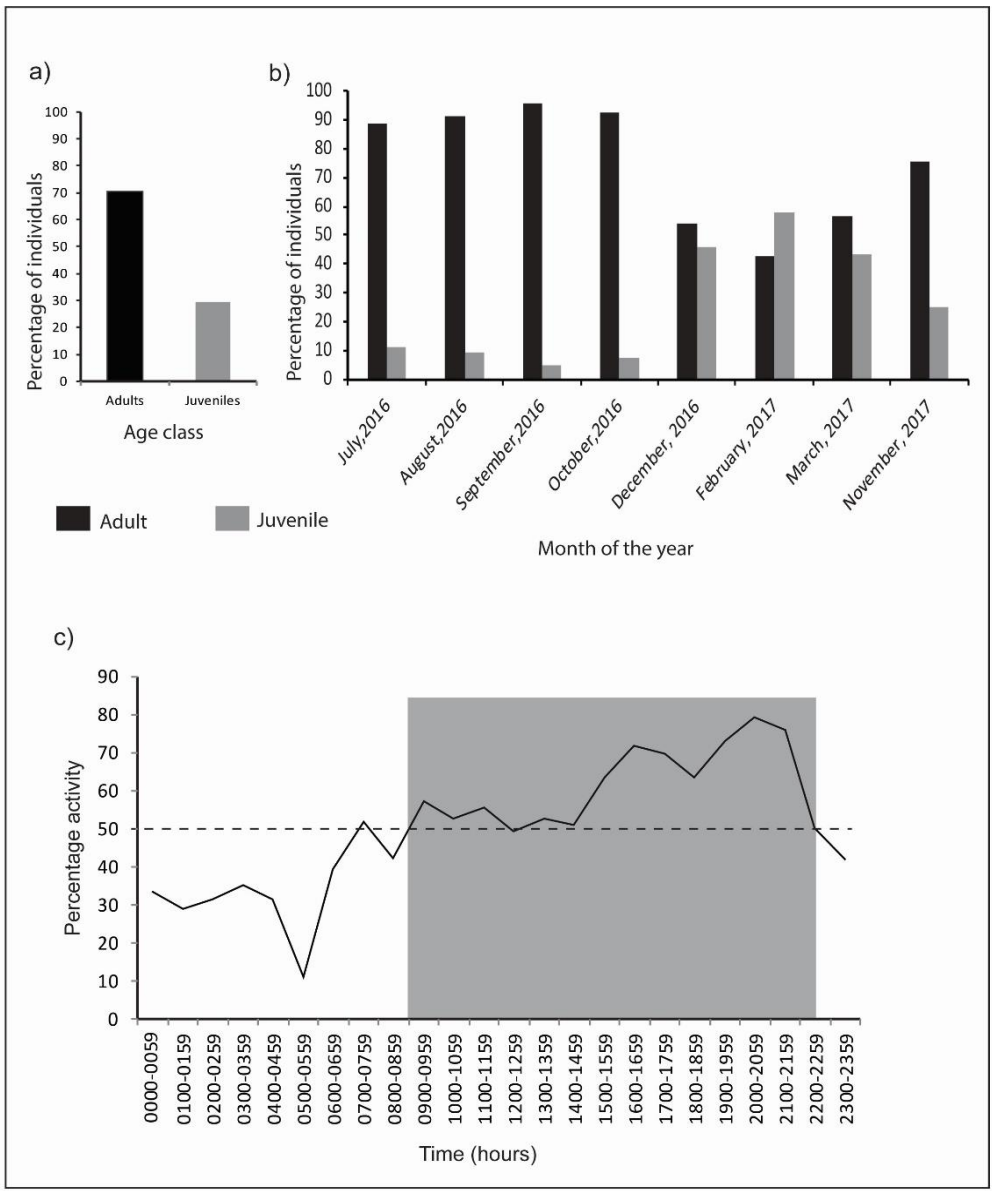

**Figure 2**: Bar graphs showing a) the percentage of adults and juveniles in the total sample; b) the percentage of adults (black bars) and juveniles (gray bars) sighted in each month of data collection (See Supplementary information for details). Comparisons are between age classes in each month of data collection; different letters signify the statistical difference. Activity levels appeared to vary widely across the 24 hours of the day. c) Line graph showing activity levels of dogs across 24 hours of the day. The highlighted portion shows the window of more than 50% activity.

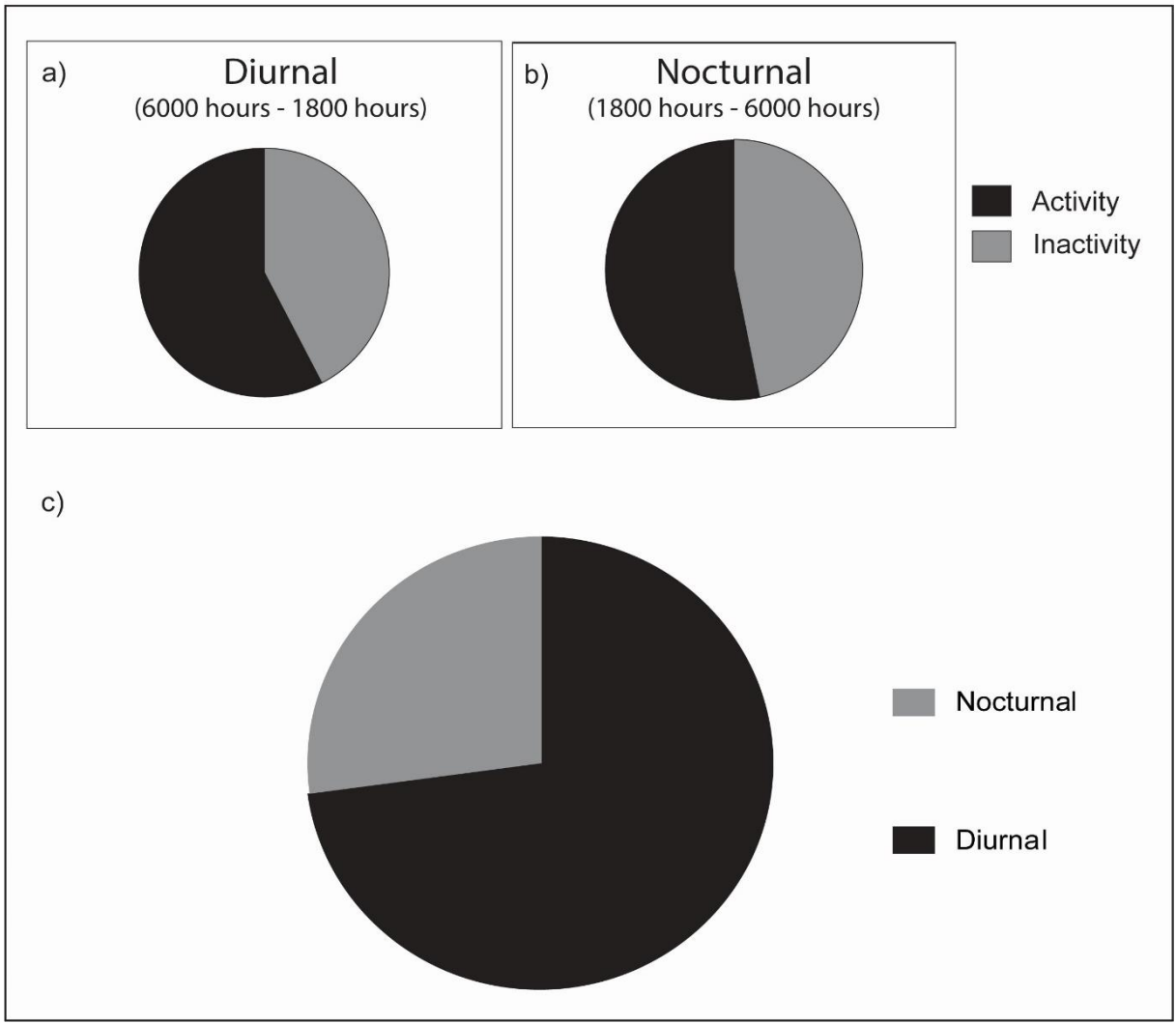

**Figure 3:** Pie-charts showing comparisons between the levels of activity and inactivity for a) diurnal (0600-1800h), b) nocturnal (1800-0600h) hours of the day and c) comparison of activity levels in the diurnal and nocturnal hours of the day.

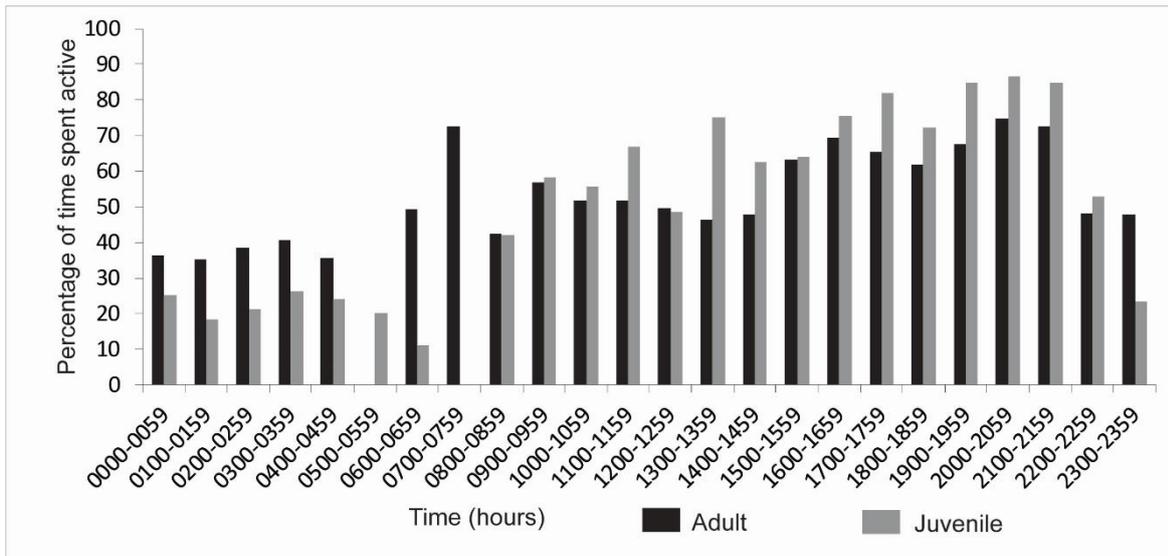

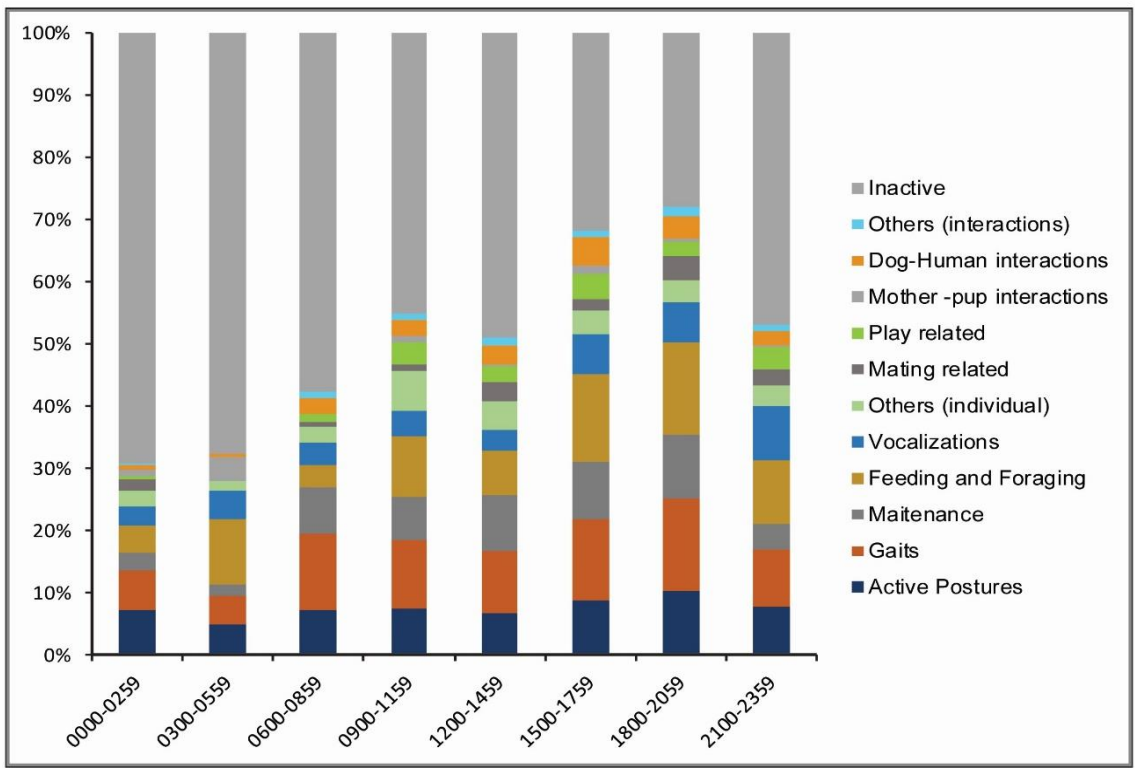

**Figure 4:** a) A bar graph showing a comparison of activity levels of adults and juveniles over 24 hours of the day. b) A stacked bar graph showing proportions of different behaviours observed within activities across 24 hours of the day.

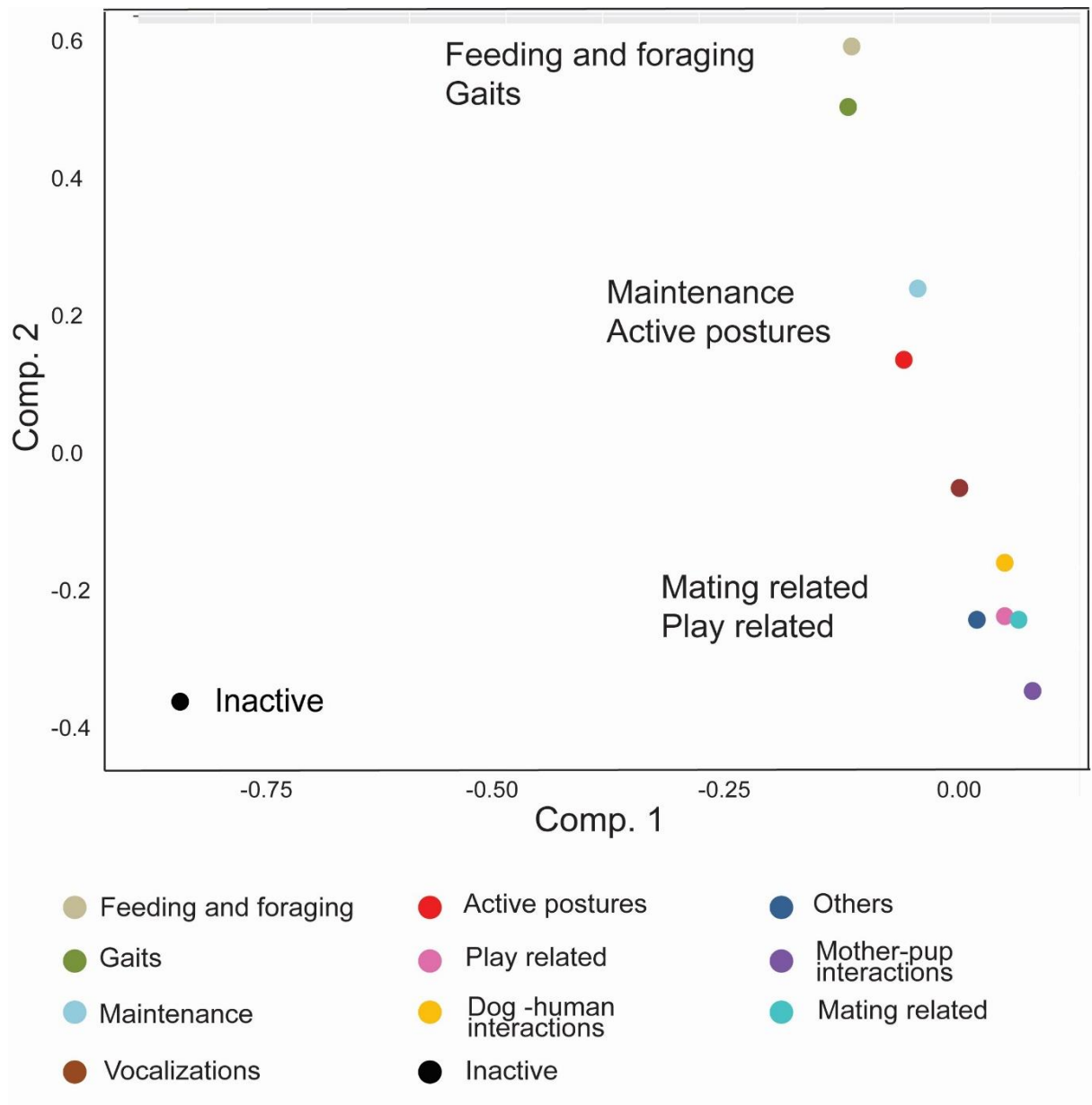

**Figure 5:** A scatter plot showing results from the PCA. Component 1 accounts for energy expenditure in behaviours and Component 2 accounts for the time of occurrence.

**Table**

| Fixed Effects: | | | | |
|---|---|---|---|---|
| | Estimate | Std. Error | z value | Pr(>|z|) |
| Age_Class | 0.511412 | 0.159933 | 3.198 | <0.001 |
| Time | -0.056343 | 0.006379 | -8.833 | <<0.001 |
| Age_Class:Time | -0.043415 | 0.011309 | -3.839 | <0.001 |
| Random Effects: | | | | |
| Groups | | Variance | | Std. Dev. |
| Locality | | 0.0131 | | 0.1146 |
| Date | | 0.08895 | | 0.2982 |

**Table 1: Table summarising the results from the GLMM analysis.**